\documentclass[12pt]{iopart}
\usepackage{iopams}
\usepackage{lmodern}
\usepackage[T1]{fontenc}
\usepackage{bbold}
\usepackage{verbatim} 
\usepackage{multicol}
\usepackage{graphicx}
\usepackage{cleveref}
\usepackage{anysize} 
\marginsize{1.5cm}{1cm}{1cm}{1cm}
\graphicspath{{graphics/}}
\usepackage[super]{nth}
\usepackage{float}  
\usepackage{cite}

\usepackage{xcolor}
\usepackage{soul}

\begin{document}

\title{Search of stochastically gated targets with diffusive particles under resetting}
\author{Gabriel Mercado-V\'asquez\footnote{\texttt gabrielmv.fisica@gmail.com} \ and Denis Boyer\footnote{\texttt boyer@fisica.unam.mx}\\
{\normalsize\it Instituto de F\'isica, Universidad Nacional Aut\'onoma de M\'exico, Mexico City 04510, Mexico}}



\begin{abstract}
    The effects of Poissonian resetting at a constant rate $r$ on the reaction time between a Brownian particle and a stochastically gated target are studied. The target switches between a reactive state and a non-reactive one. We calculate the mean time at which the particle subject to resetting hits the target for the first time, while the latter is in the reactive state. The search time is minimum at an optimal resetting rate that depends on the target transition rates.  When the target relaxation rate is much larger than both the resetting rate and the inverse diffusion time, the system becomes equivalent to a partially absorbing boundary problem. In other cases, however, the optimal resetting rate can be a non-monotonic function of the target rates, a feature not observed in partial absorption.
    We compute the relative fluctuations of the first hitting time around its mean and compare our results with the ungated case. The usual universal behavior of these fluctuations for resetting processes at their optimum breaks down due to the target internal dynamics.

\end{abstract}

\section{Introduction}

A reactant in a physicochemical  system is said to be gated when it switches to multiple conformational states, which alter its capacity to react with other compounds. The gating process could be due to both fluctuations in the environment and internal mechanisms of the reactants. The simplest gating process is the two-state model in which a reactant transits back and forth between an open state, that represents a
reactive conformation, and a closed, non-reactive state.  

There is a variety of examples in which reactions between the compounds of a system are controlled by changes in their conformational states, ranging from natural processes such as protein binding\cite{mccammon1981gated,szabo1982stochastically,spouge1996,zhou1996theory,Berezhkovskii1997}, gene expression \cite{mcadams1997stochastic,tian2006stochastic,chubb2010bursts,suter2011mammalian,munsky2012,wu1997chromatin,eberharter2002histone} or cellular transport mediated by ion-channels\cite{bressloff2014stochastic,reingruber2009gated,sakmann2013single}, to artificial processes such as the diffusion of particles in synthetic nanopores\cite{xia2008gating}, or more general intermittent search processes\cite{Benichou2011}. Whether we are interested in knowing the rate at which two proteins bind to each other or in calculating the flux of ions across a gating channel in the cell membrane, the problem can be often reduced to the generic one of computing the time at which a diffusive particle reaches for the first time a target site in its reactive state.

A two-state model with diffusive particles was first studied in the pioneering work of McCammon and Northrup \cite{mccammon1981gated}. In this work, the authors computed the association rate for the case where the non-reactive
periods were sufficiently long. Shortly after, more complex systems in which particles could transit between several conformational states were analyzed\cite{szabo1982stochastically,spouge1996,zhou1996theory,Berezhkovskii1997}. Recently, the topic of gated reactions has recovered interest and has been retaken not only for the problem of a Brownian particle on the infinite line\cite{PRLFirstHittingTimes}, but also in other contexts such as in random walks on networks\cite{scher2021unified}, run-and-tumble motion\cite{mercado2021first} and diffusion with stochastic resetting in an interval\cite{Bressloff_2020}. 

Due to the interplay between the kinetics of the system compounds and the gating process, it is clear that the  motion of diffusing entities strongly affects the reaction time. In the context of perfectly reactive targets, non-Brownian search processes have recently attracted attention as they may significantly reduce reaction times. Among such processes, diffusion under stochastic resetting have received a lot of attention. As shown in the seminal work of Evans and Majumdar \cite{Evans2011PRL}, stochastic resetting can expedite the mean time needed by a Brownian particle to be absorbed on a fixed target site.

In this original model, the resetting process consists in randomly interrupting particle diffusion on the infinite line at some constant rate  and bringing it back to a fixed position, from which the diffusion process starts anew. Resetting the particle motion has important consequences on the first passage properties\cite{evans2020stochastic,Evans_2018}. The mean first passage time (MFPT) at the absorbing target becomes finite and can be minimized with respect to the resetting rate\cite{Evans2011PRL,evans2011diffusion}. Research on resetting processes has further unveiled that a similar optimization can be achieved in a variety of situations, such as diffusion with time-dependent resetting rates\cite{pal2016diffusion,nagar2016diffusion}, other non-Poissonian resetting protocols\cite{ChechkinPRL2018,Eule_2016,MonteroPRE2016}, resetting with refractory periods\cite{evans2018effects}, resetting in bounded domains\cite{Christou_2015} or involving anomalous diffusion processes\cite{Kusmierz2014PRL,KusmierzPRE2015,KusmierzPRE2019,MasoPRE2019}, to name a few (see \cite{evans2020stochastic} for a review).
Moreover, the optimization by stochastic resetting is not exclusive to the searches of simple targets, \textit{i.e.}, targets that are perfectly reactive, but has also been studied in the case of partially absorbing targets \cite{WhitehousePartialPRE2013,schumm2021search} and for stochastically gated targets\cite{Bressloff_2020}. 

Among the distinctive features of stochastic resetting, such as the emergence of non-equilibrium steady states \cite{manrubia1999stochastic,evans2011diffusion,Evans_2018} and their peculiar relaxation dynamics \cite{majumdar2015dynamical,Gupta_2021,Singh_2020}, one should mention the universal behaviour of the relative standard deviation of the first passage time distribution, which becomes unity at optimality (when there exists a finite optimal resetting rate) \cite{ReuveniPRLoptimal,PalPRL2017,BelanPRL2018}. Notably, this result is valid for all types of search dynamics, even if the search process in the absence of resetting has an infinite MFPT. As we will illustrate further, this feature no longer holds when the target follows its own dynamics independently of the resetting process.

In the present work, we study the first hitting statistics between a particle, which stochastically resets to its initial position on the semi-infinite line, and a gated target that intermittently switches between two states: a reactive state that absorbs the diffusive particle upon encounter, and a non-reactive one which reflects the particle. We calculate the survival probabilities of the particle at time $t$, and further deduce quantities of interest such as the first two moments of the hitting time distribution. As is usual in resetting processes, the mean first hitting time (MFHT) can be optimized by a suitable choice of the resetting rate. We study the behaviour of the optimal resetting rate as a function of the target dynamical parameters. From this analysis emerges a strong connection between our model and the problem of diffusion with stochastic resetting in the presence of a partially absorbing target. We show how the two problems actually become equivalent in the limit of high transition rates, or when the target is in the non-reactive state most of the time. We also analyse the relative variance of the first hitting time around the mean and study its dependence with respect to the target rates and  the resetting rate. The relative fluctuations are no longer unity at the optimal resetting rate, and can take much larger values instead. This is due to the fact that the dynamics of the target state is independent of the resetting process itself. The problem therefore differs from the one considered in \cite{Bressloff_2020}, where the search of a gated target by diffusion under resetting was studied through a renewal approach, that assumed that the resetting process also acted on the target state.

The paper is organized as follows: we begin in Section \ref{Setup} by introducing the model and deduce the equations of motion that govern the survival probabilities, which are solved in the Laplace space. With these solutions, in Section \ref{secMFHT} we find an exact expression for the MFHT and analyze its behaviour as a function of the target transition rates and of the resetting rate. In Section \ref{secPartial} we discuss the connection between our model and the partial absorption problem. Section \ref{secCav} is devoted to the analysis of the relative variance of the first hitting times, and we conclude in Section \ref{secDiscussion}.  A comparison between our findings and those of \cite{Bressloff_2020} is discussed in more details in \ref{secBressloff}. 

\section{The problem and its solution}\label{Setup}

\begin{figure}[ht]
    \centering
    \includegraphics[width=.5\textwidth]{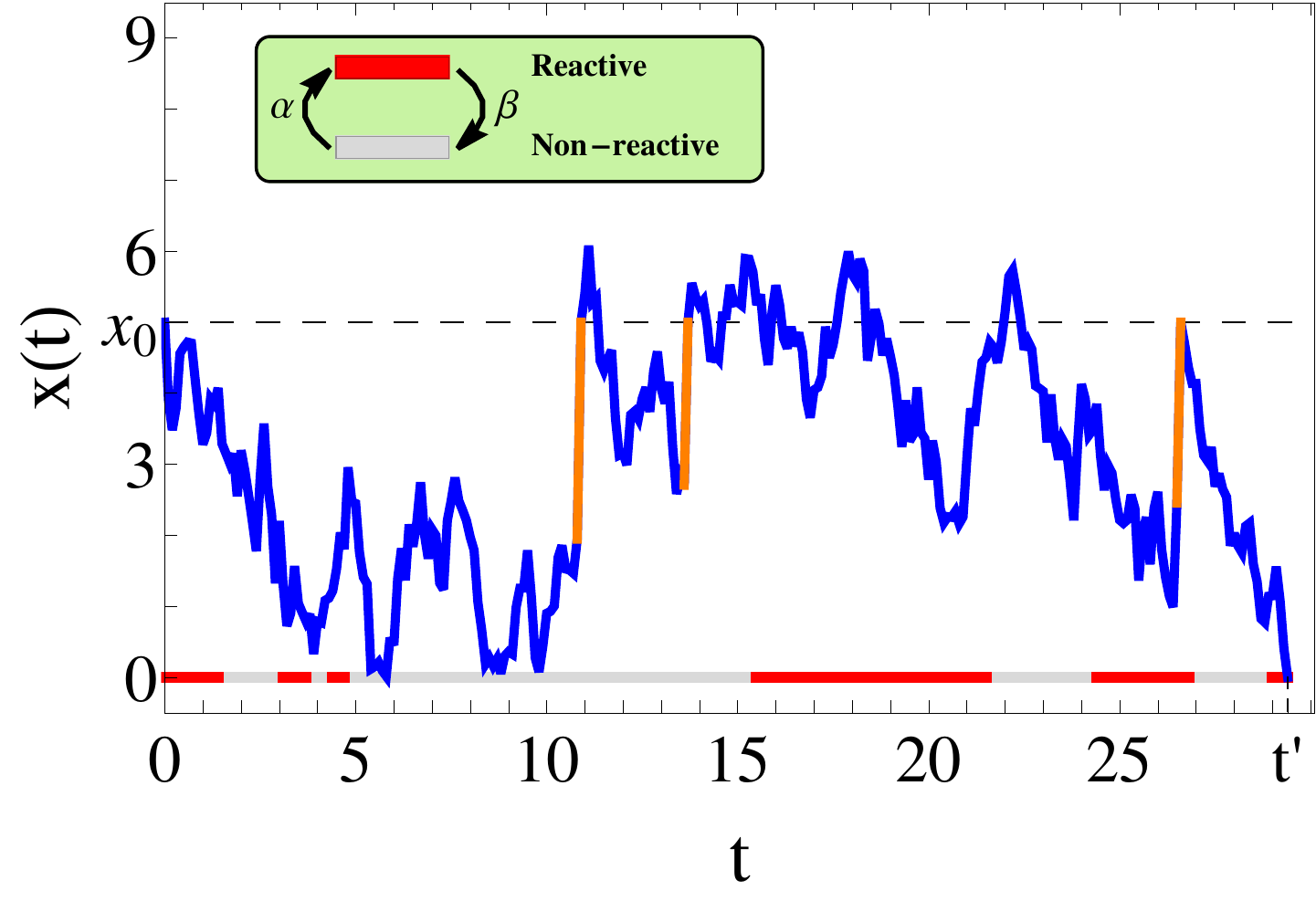}
    \caption{Trajectory of a diffusive particle (blue line) in 1D, in the presence of an intermittent target placed at the origin. The periods of time during which the target is reactive (or absorbing) are represented by red segments, whereas the gray intervals represent the target in the non-reactive (or reflective) state. At exponentially distributed time intervals with mean $1/r$, the particle is reset to the position $x_r$ (orange line),  which coincides in this example with the initial position $x_0$.}
    \label{fig:1}
\end{figure}

Let us consider on the semi-infinite line a Brownian particle with diffusion coefficient $D$, starting at $t=0$ from a position $x_0>0$, and which is subject to a stochastic Poissonian resetting process of rate $r$. The resetting position is denoted as $x_r>0$. At the origin, a stochastically gated target is placed. The dynamics of the target will be characterized by the time-dependent binary variable $\sigma(t)$, which takes the value $\sigma=0$ when the target is non-reactive, and $\sigma=1$ when it is reactive. The target stochastically switches from the state $0$ to $1$ with rate $\alpha$, whereas it switches from the state $1$ to $0$ with rate $\beta$ (see Fig. \ref{fig:1}). The diffusing particle is absorbed upon its first encounter with the target in the reactive state.

We define $Q_0(x_0,t)$ as the probability that the particle has not hit the target up to time $t$, given the initial position $x_0$ and initial target state $\sigma(t=0)=0$ [the variable $x_r$ is implicit]. Similarly, we define $Q_1(x_0,t)$ for the initial target state $\sigma(t=0)=1$. In \ref{appendixA} we show that these probabilities satisfy the coupled backward Fokker-Planck equations
\begin{eqnarray}
    \frac{\partial Q_0 (x_0,t)}{\partial t}=D\frac{\partial^2 Q_0 (x_0,t)}{\partial x_0^2}+\alpha(Q_1(x_0,t)-Q_0(x_0,t))+r(Q_0 (x_r,t)-Q_0 (x_0,t)),\label{BFPQ0}\\
     \frac{\partial Q_1 (x_0,t)}{\partial t}=D\frac{\partial^2 Q_1 (x_0,t)}{\partial x_0^2}+\beta(Q_0(x_0,t)-Q_1(x_0,t))+r(Q_1(x_r,t)-Q_1(x_0,t)).\label{BFPQ1}
\end{eqnarray}
The system of equations (\ref{BFPQ0}) and (\ref{BFPQ1}) will satisfy the following boundary conditions:
\begin{eqnarray}
Q_1(x_0=0,t)=0,\label{conditionQ}\\
\frac{\partial Q_0(x_0,t)}{\partial x_0}\Big|_{x_0=0}=0.\label{conditionQzero}
\end{eqnarray}
Eq.  (\ref{conditionQ})   enforces the absorbing condition of the target in the reactive state, whereas Eq.  (\ref{conditionQzero}) asserts that the target in the non-reactive state will reflect the diffusive particle upon encounter (see \cite{PRLFirstHittingTimes} for a detailed derivation of the latter condition).

We also define the average survival probability $Q_{av}(x_0,t)$ for the particle starting at $x_0$ that results from averaging over the initial target states generated
by the steady-state distribution of the two-state Markov chain:
\begin{equation}\label{qav}
Q_{av}(x_0,t)=\frac{\beta}{\alpha+\beta}Q_0(x_0,t)+\frac{\alpha}{\alpha+\beta}Q_1(x_0,t).
\end{equation}


The probability distributions of the first hitting time $t$ are denoted as $P_0(x_0,t)$ and  $P_1(x_0,t)$, with the same notations as before for the initial conditions. These first hitting time densities (FHTDs) are deduced from the survival probabilities through the usual relation\cite{redner2001guide}:
\begin{equation}
P_{0,1}(x_0,t)=-\frac{\partial Q_{0,1}(x_0,t)}{\partial t}.\label{fhtd}
\end{equation}

Introducing the Laplace transforms $\widetilde{Q}_{0,1}(x_0,s)=\int^\infty_0e^{-st}Q_{0,1}(x_0,t)dt$ and using the initial condition $Q_{0,1}(x_0,t=0)=1$ for $x_0>0$, Eqs. (\ref{BFPQ0}) and (\ref{BFPQ1}) become 
\begin{eqnarray}
D\frac{\partial^2 \widetilde{Q}_0 (x_0,s)}{\partial x_0^2}+\alpha \widetilde{Q}_1(x_0,s)-(s+\alpha+r)\widetilde{Q}_0(x_0,s)=-1-r\widetilde{Q}_0 (x_r,s),\label{syst1}\\
D\frac{\partial^2 \widetilde{Q}_1 (x_0,s)}{\partial x_0^2}+\beta \widetilde{Q}_0(x_0,s)-(s+\beta+r)\widetilde{Q}_1(x_0,s)=-1-r\widetilde{Q}_1 (x_r,s),
\label{syst2}
\end{eqnarray} 
and the boundary conditions (\ref{conditionQ}) and (\ref{conditionQzero}) read
\begin{eqnarray}
\widetilde{Q}_1(x_0=0,s)=0,\label{conditionQLaplace}\\
\frac{\partial \widetilde{Q}_0(x_0,s)}{\partial x_0}\Big|_{x_0=0}=0.\label{conditionQzeroLaplace}
\end{eqnarray}
By using Eq. (\ref{fhtd}) and integrating by parts, the Laplace transform of the FHTD will be simply given by 
\begin{equation}
    \widetilde{P}_{0}(x_0,s)=1-s\widetilde{Q}_{0}(x_0,s)\ {\rm and}\   \widetilde{P}_{1}(x_0,s)=1-s\widetilde{Q}_{1}(x_0,s).\label{relQtoP}
\end{equation}
We consider $Q_0(x_r,s)$ and $Q_1(x_r,s)$ as unknown inhomogeneous terms in the differential equations (\ref{syst1}) and (\ref{syst2}). The homogeneous part of this system is solved with the ansatz $\boldsymbol{\xi} e^{\lambda x_0}$, where the vector $\boldsymbol{\xi}$ and $\lambda$ are determined from solving
\begin{equation}
    \left(\begin{array}{cc} 
    D\lambda^2-(\alpha+r+s) & \alpha \\
	 \beta & D\lambda^2-(\beta+r+s)
    \end{array}\right)\boldsymbol{\xi}=0.\label{matrixsigma}
\end{equation}
After straightforward algebra, the general solution $\mathbf{\widetilde{Q}}= \left(\begin{array}{cc}
\widetilde{Q}_0 & \widetilde{Q}_1 \end{array}\right)^T$ is given by the following linear combination
\begin{equation}
\mathbf{\widetilde{Q}}=A_1\boldsymbol{\xi}_1e^{-\lambda_1 x_0}+A_2\boldsymbol{\xi}_1e^{\lambda_1 x_0}+A_3\boldsymbol{\xi}_2e^{-\lambda_2 x_0}+A_4\boldsymbol{\xi}_2e^{\lambda_2 x_0}+\mathbf{\widetilde{Q}}^{inh},\label{gralsol}    
\end{equation} 
where $\mathbf{\widetilde{Q}}^{inh}=\left(\begin{array}{cc}
\widetilde{Q}_0^{inh} & \widetilde{Q}_1^{inh}\end{array}\right)^T$ is the constant solution given by
\begin{eqnarray}
\widetilde{Q}_0^{inh}=\frac{1+r\widetilde{Q}_0(x_r,s)}{D\lambda^2_1}+\frac{r\alpha\left[ \widetilde{Q}_1(x_r,s)-\widetilde{Q}_0(x_r,s)\right]}{D\lambda^2_1\lambda^2_2},\label{Qinh0}\\
\widetilde{Q}_1^{inh}=\frac{1+r\widetilde{Q}_1(x_r,s)}{D\lambda^2_1}+\frac{r\beta\left[\widetilde{Q}_0(x_r,s)-\widetilde{Q}_1(x_r,s)\right]}{D^2\lambda^2_1\lambda^2_2}.\label{Qinh1}
\end{eqnarray}
The factors $A_k$ are determined from the boundary conditions and the no-divergence of the probabilities $Q_{0,1}$ as $x_0\to\infty$. The roots $\lambda_1$ and $\lambda_2$ in Eqs. (\ref{gralsol})--(\ref{Qinh1}) are given from (\ref{matrixsigma}) by
\begin{equation}
\lambda_1=\sqrt{\frac{s+r}{D}}, \quad \lambda_2=\sqrt{\frac{s+\alpha+\beta+r}{D}},
\end{equation}
whereas the vectors $\boldsymbol{\xi}_1$ and $\boldsymbol{\xi}_2$ are
\begin{equation*}
	\boldsymbol{\xi}_1=\left(\begin{array}{c} 
	1\\
	1\end{array}\right)
	, \quad \boldsymbol{\xi}_{2}=\left(\begin{array}{c} 
	-\frac{\alpha}{\beta}\\
	1\end{array}\right).
\end{equation*}
To avoid infinite solutions at $x_0\to\infty$, we must set $A_2=A_4=0$ in Eq. (\ref{gralsol}). From the boundary conditions (\ref{conditionQLaplace})-(\ref{conditionQzeroLaplace}) we obtain the remaining constants,
\begin{eqnarray}
A_1=-\frac{\alpha\lambda_2\widetilde{Q}_1^{inh}}{\alpha\lambda_2+\beta\lambda_1},
\end{eqnarray}
and $A_3=\frac{\beta\lambda_1}{\alpha\lambda_2}A_1$. Substituting these factors into Eq.  (\ref{gralsol}), 
\begin{eqnarray}
    \widetilde{Q}_0(x_0,s)&=-\frac{\alpha\lambda_2}{\alpha\lambda_2+\beta\lambda_1}\left(e^{-\lambda_1 x_0}-\frac{\lambda_1}{\lambda_2}e^{-\lambda_2 x_0}\right)\widetilde{Q}_1^{inh}+\widetilde{Q}_0^{inh},\label{Q0andQinh}\\
    \widetilde{Q}_1(x_0,s)&=-\frac{\alpha\lambda_2}{\alpha\lambda_2+\beta\lambda_1}\left(e^{-\lambda_1 x_0}+\frac{\beta\lambda_1}{\alpha\lambda_2}e^{-\lambda_2 x_0}\right)\widetilde{Q}_1^{inh}+\widetilde{Q}_1^{inh}.\label{Q1andQinh}
\end{eqnarray}
The average survival probability takes a slightly simpler form:
\begin{equation}
    \widetilde{Q}_{av}(x_0,s)=-\frac{\alpha\lambda_2\widetilde{Q}_1^{inh}}{\alpha\lambda_2+\beta\lambda_1}e^{-\lambda_1 x}+\frac{1+r \widetilde{Q}_{av}(x_r,s)}{D\lambda^2_1}.\label{Qaverage}
\end{equation}

Substituting Eqs. (\ref{Qinh0})-(\ref{Qinh1}) into Eqs. (\ref{Q0andQinh})-(\ref{Q1andQinh}), and then setting $x_r=x_0$, one obtains in a self-consistent way the survival probabilities $\tilde{Q}_0(x_0,s)$ and $\tilde{Q}_1(x_0,s)$, {\it i.e.}, when the initial position is the resetting position:
\begin{eqnarray}
    \widetilde{Q}_0(x_0,s)&=\frac{\alpha  \lambda _2 \left(e^{\lambda _1 x_0}-1\right)+\lambda _1 \left(\beta +(\alpha +r) e^{-\lambda _2 x_0}\right)e^{\lambda _1 x_0}-\frac{s\lambda _1 r }{\alpha +\beta +s}e^{(\lambda_1-\lambda _2)x_0}}{ \alpha  \lambda _2r +se^{\lambda _1 x_0}\left[\left(\beta  \lambda _1+\alpha  \lambda _2\right) +\frac{\beta  \lambda _1 r }{\alpha +\beta +s}e^{-\lambda _2 x_0}\right]},\label{Q0x0}\\
    \widetilde{Q}_1(x_0,s)&=\frac{ \alpha  \lambda _2 \left(e^{\lambda _1 x_0}-1\right)+\beta\lambda _1   \left(1-e^{-\lambda _2 x_0}\right)e^{\lambda _1 x_0}}{ \alpha  \lambda _2r+se^{\lambda _1 x_0}\left[\left(\beta  \lambda _1+\alpha  \lambda _2\right) +\frac{\beta  \lambda _1 r }{\alpha +\beta +s}e^{-\lambda _2 x_0}\right]}.\label{Q1x0}
\end{eqnarray}
whereas the average survival probability is
\begin{equation}
    \widetilde{Q}_{av}(x_0,s)=\frac{\alpha  \lambda _2 \left(1-e^{-\lambda_1 x_0}\right) (s+\alpha +\beta )+\beta  \lambda _1 \left(r e^{-\lambda_2 x_0}+s+\alpha +\beta \right)}{\alpha \lambda _2 \left(r e^{-\lambda _1x_0}+s\right)(s+\alpha +\beta )+s\beta  \lambda _1 \left(r e^{-\lambda_2x_0}+s+\alpha +\beta \right)}.\label{Qavx0}
\end{equation}


\section{Mean first hitting time}\label{secMFHT}

 In the following we keep considering $x_r=x_0$ (resetting to the starting position) and define the mean first hitting time given the initial target condition $\sigma=0$ ($\sigma=1$, respectively) as $T_0(x_0)$ ($T_1(x_0)$, respectively). These quantities are obtained from the usual relation $T_{0,1}(x_0)=\int^\infty_0 Q_{0,1}(x_0,t)dt=\widetilde{Q}_{0,1}(x_0,s=0)$. Setting $s=0$ in Eqs. (\ref{Q0x0}) and (\ref{Q1x0}), one deduces
\begin{eqnarray}
T_0(x_0)=\frac{e^{\sqrt{\frac{r}{D}}x_0}-1}{r}+\frac{\beta+(r+\alpha)e^{-\sqrt{\frac{r+\alpha+\beta}{D}}x_0}}{\alpha\sqrt{r(r+\alpha+\beta)}}e^{\sqrt{\frac{r}{D}}x_0},\label{T0tot}\\
T_1(x_0)=\frac{e^{\sqrt{\frac{r}{D}}x_0}-1}{r}+\frac{\beta}{\alpha}\left(\frac{1-e^{-\sqrt{\frac{r+\alpha+\beta}{D}}x_0}}{\sqrt{r(r+\alpha+\beta)}}\right)e^{\sqrt{\frac{r}{D}}x_0}.
\label{T1tot}\end{eqnarray}
From Eq.  (\ref{Qavx0}), the average mean first hitting time reads
\begin{equation}
    T_{av}(x_0)=\frac{e^{\sqrt{\frac{r}{D}}x_0}-1}{r}+\frac{\beta e^{\sqrt{\frac{r}{D}}x_0}(\alpha+\beta+re^{-\sqrt{\frac{r+\alpha+\beta}{D}}x_0})}{\alpha(\alpha+\beta)\sqrt{r(r+\alpha+\beta)}}.\label{mfptTav}
\end{equation}
As well-known for the case of perfectly absorbing targets \cite{Evans2011PRL,evans2011diffusion}, one of the main consequence of introducing resetting in the dynamics of the diffusive particle is to make the mean of the FHTD finite, unlike in free diffusion, where it diverges. Furthermore, the different MFHTs here can be minimized by a suitable choice of the resetting rate.

The solution of the mean first hitting time of the gated problem calls for several comments. As expected, if we set $\beta=0$ in Eq. (\ref{T0tot}) or (\ref{T1tot}), we recover the expression of the MFHT for the ungated case, denoted as $T_r(x_0)$ here:
\begin{equation}
     T_{av}(x_0,\beta=0)= T_{r}(x_0)=\frac{e^{\sqrt{\frac{r}{D}}x_0}-1}{r}.\label{Tresetting}
\end{equation}
$T_r$ is a non-monotonic function of $r$ that is minimum at the optimal resetting rate $r^*(\beta=0)=2.53963...D/x_0^2$, a result first deduced in \cite{Evans2011PRL}.

\begin{figure}[htp]
    \centering
    \includegraphics[width=\textwidth]{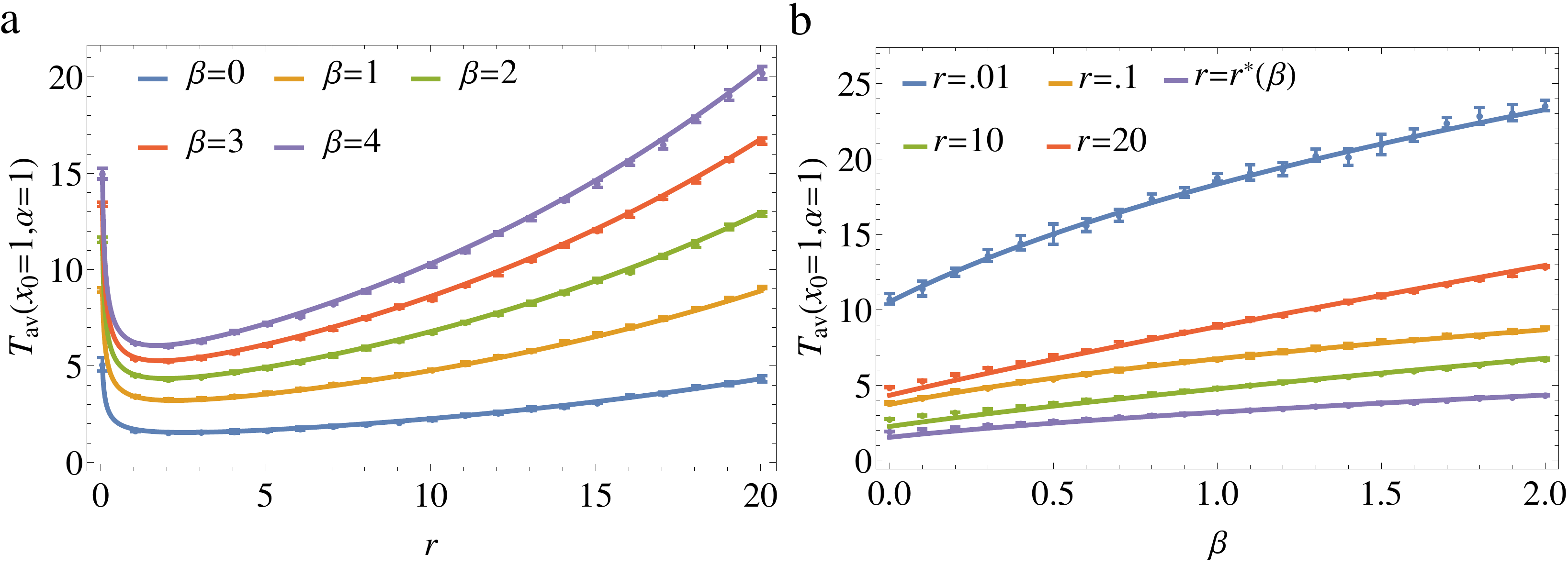}
    \caption{(a) Mean first hitting time $T_{av}$ as a function of $r$ for several values of $\beta$ ($x_0=1$, $D=1$ and $\alpha=1$). (b) Same quantity as a function of $\beta$ for several values of $r$. Symbols represent simulation results obtained with the Gillespie algorithm\cite{GILLESPIE1976403}.}
    \label{fig:2ab}
\end{figure}

The solution for the average MFHT in Eq.  (\ref{mfptTav}) also exhibits a non-monotonic behaviour with a single minimum (Fig. \ref{fig:2ab}a), for all parameter values $\alpha,\beta>0$ of the intermittent target. The optimal resetting rate $r^*$ that minimizes the MFHT varies with the switching parameters $\alpha$ and $\beta$. Increasing the parameter $\beta$ makes the target less reactive, which causes an increase of the MFHT. As shown by Fig. \ref{fig:2ab}b, at  a fixed resetting rate, the MFHT increases monotonically with $\beta$. Even when the switching parameter $\beta$ is high, an optimal resetting rate $r=r^*(\beta)$ can always be found. Therefore, fixing $\alpha$, it is possible to draw a minimal curve for the MFHT as a function of $\beta$. As depicted in Fig. \ref{fig:2ab}b, any MFHT with another value of $r$ will lie above the curve corresponding to $r^*(\beta)$. One can also notice the non-monotonic variations of the MFTH with $r$: the MFHT first decreases with $r$ until it reaches its minimal value at $r^*(\beta)$, which is of order one. For $r> r^*(\beta)$, the MFPT increases with $r$. A very good agreement with numerical simulations is obtained.


 In Eq. (\ref{mfptTav}), the dependence of the MFHT with respect to the target rates is not as simple as one would wish and obtaining an analytical expression for $r^*$ seems beyond reach. Below, we derive a simplified expression in the limiting case when the target rapidly switches between the reactive and non-reactive states, and compare the results with the numerical minimization of the exact solution (\ref{mfptTav}).

\begin{figure}[htp]
    \centering
    \includegraphics[width=\textwidth]{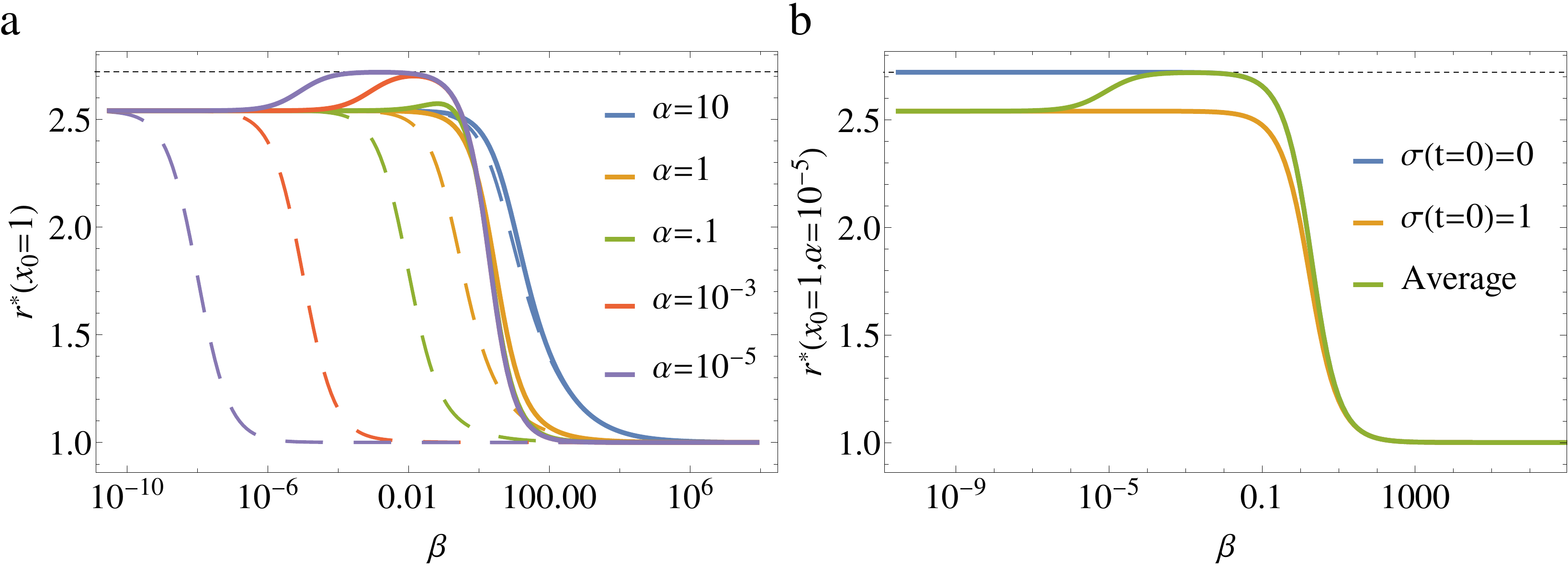}
    \caption{ a) Optimal resetting rate $r^*$ as a function of $\beta$ and several values of $\alpha$ (fixing $x_0=1$ and $D=1$). The continuous lines are obtained from numerical minimization of the exact result  (\ref{mfptTav}), whereas the dashed lines represent the solution of the approximate Eq. (\ref{optBetaLarge}). b) Optimal resetting rates obtained from the minimizations of the functions $T_0$, $T_1$ and $T_{av}$, respectively, see Eqs. (\ref{T0tot}), (\ref{T1tot}) and (\ref{mfptTav}), for a small value of $\alpha$ ($10^{-5}$). In both figures, the upper horizontal dotted line represents the maximum value $2.72033...D/x_0^2$ that $r^*$ can reach, which is given by minimizing Eq. (\ref{T0smallalpha}).}
    \label{fig:3}
\end{figure}

In the limit of large $\alpha$ and $\beta$ compared to $r$, we approximate $\sqrt{r+\alpha+\beta}\approx\sqrt{\alpha+\beta}$ in Eq.  (\ref{mfptTav}) and can always neglect the term proportional to $e^{-x_0\sqrt{\frac{r+\alpha+\beta}{D}}}$ to obtain
\begin{equation}
    T_{av}(x_0)\approx\frac{e^{\sqrt{\frac{r}{D}}x_0}-1}{r}+\frac{\beta e^{\sqrt{\frac{r}{D}}x_0}}{\alpha\sqrt{r(\alpha+\beta)}}.\label{mfptbeta}
\end{equation}
Defining the dimensionless parameters
\begin{eqnarray}
    z&=x_0\sqrt{\frac{r}{D}},\label{z}\\
    w&=\frac{\beta\sqrt{r}}{2\alpha\sqrt{\alpha+\beta}},
\end{eqnarray}
the approximate optimal resetting rate obeys the transcendental equation
\begin{equation}
    \frac{z}{2}-1+e^{-z}+w(z-1)=0.\label{optBetaLarge}
\end{equation}
The solution of Eq.  (\ref{optBetaLarge}) as a function of $\beta$ is shown in Fig. \ref{fig:3}a (dashed lines), together with the exact optimal parameter obtained from numerical minimization of Eq.  (\ref{mfptTav}). Clearly, the two solutions show a good agreement for all $\beta$ only if  $\alpha \gg r^*$. Otherwise, the differences are significant in the intermediate regime of $\beta$. 

In the high transition rates regime, if the target is mostly non-reactive  ($r\ll\alpha\ll\beta$, such that $w\gg 1$), the first three terms of the left hand side of Eq.  (\ref{optBetaLarge}) can be neglected and we arrive at the simple solution $z=1$. 
From Eq. (\ref{z}),  the optimal resetting rate in the limit $\beta=\infty$ is therefore given by
\begin{equation}
    r^*(\beta=\infty)=D/x^2_0,\label{roptlargebeta}
\end{equation}
which is substantially lower than the optimal rate $r^*(\beta=0)=2.53963...D/x_0^2$ for the ungated target (see Fig. \ref{fig:3}a). Therefore, to optimize the search process of a poorly reactive target, one must opt for less frequent resetting compared with the perfectly reactive case, at a rate exactly given by the inverse diffusion time $D/x_0^2$.
It is also worth noting that, even though the expression (\ref{optBetaLarge}) is obtained in the high transition rates limit, we can recover the solution for the ungated case: setting $\beta=0$, it reduces to the transcendental equation $\frac{z^*}{2}-1+e^{-z^*}=0$, whose solution is $z^*=1.59362...$ or $r^*(\beta=0)=2.53963...D/x_0^2$. 

As shown by Fig. \ref{fig:3}a, $r^*$ always remains of the order of the inverse diffusion time $D/x_0^2$. Nevertheless, the (exact) optimal resetting rate does not always decrease as the target becomes less reactive.  At odds with the solution given by Eq. (\ref{optBetaLarge}), $r^*$ can exhibit a clear non-monotonic shape with respect to $\beta$, with a maximum at a value above $2.53963...D/x_0^2$.
This occurs when the parameter $\alpha$ is fixed to a small value (compared to the inverse diffusion time), a regime where the approximation (\ref{mfptbeta}) is no longer valid.
In this case, $r^*$ is maximum for a value of $\beta$ which is larger than $\alpha$, namely, in a situation where the target is most of the time inactive. 

This non-monotonic shape of the optimal resetting rate stems from properties exhibited by the two MFHTs $T_0$ and $T_1$. As depicted in Fig. \ref{fig:3}b, the resetting rates that minimize $T_0$ and $T_1$ taken separately are different. If the target is initially reactive, it remains so during a random time of mean $1/\beta$ until it switches to the non-reactive state. For a small transition rate $\beta$, the initial reactive phase can thus be very long and the target is considered as practically ungated: $r^*$ coincides with the optimal rate $r^*(\beta=0)=2.53963...D/x_0^2$.  On the contrary, if the target is initially non-reactive, the searcher will diffuse and reset without being absorbed during a random time of mean $1/\alpha$ until the target becomes reactive. If this first transition happens after a long time ($\alpha$ small), the searcher will have a random position approximately distributed along the non-equilibrium steady state in the presence of the reflecting boundary. If in addition the transition rate $\beta$ is small, once the target activates, it can be considered as practically ungated and the problem becomes analogous to the standard one, but with a distribution of starting positions. As a consequence, the value of the optimal resetting rate is larger, as shown in Fig. \ref{fig:3}b (see also Eq. (\ref{T0smallalpha}) below). 

Since $T_{av}$ represents the average over the initial target states in Eq. (\ref{qav}), for values of $\beta$ much smaller than $\alpha$, the target is likely to be initially reactive, and the main contribution to $T_{av}$ comes from $T_1$. Conversely, when $\beta$ becomes greater than $\alpha$ (but still $\ll r^*$), the contribution of $T_0$ is dominant.  Therefore the resetting rate that minimizes  $T_{av}$ increases and reaches the value that minimizes $T_0$. Eventually, in the regime $\beta\gg r^*$ the resetting rate drops to the value $D/x_0^2$ discussed previously. These considerations explain the non-monotonic behaviour of $r^*$ at small $\alpha$ seen in Figs. \ref{fig:3}a-b.

The upper bound reached by the optimal resetting rate $r^*$ in our problem can be calculated as follows. With $\beta=0$, the value of $r$ that minimize $T_0$ becomes independent of $\alpha$ at small $\alpha$. This can be noticed by setting $\beta=0$ and expanding Eq. (\ref{T0tot}) around $\alpha=0$:
\begin{equation}
T_0(x_0,\alpha,\beta=0)\approx \frac{e^{\sqrt{\frac{r}{D}}x_0}-1}{r}+\frac{1-\sqrt{\frac{r}{D}}x_0}{2r}+\frac{1}{\alpha}+\mathcal{O}(\alpha).\label{T0smallalpha}
\end{equation}
In the limit $\alpha\to0$, all the terms of order $\alpha$ or higher can be neglected. Therefore, the minimization of Eq. (\ref{T0smallalpha}) with respect to $r$ will only involve the first two terms of the right hand side, leading to an optimal resetting rate of $2.72033...D/x_0^2$, independent of $\alpha$. This is the maximum value that the optimal resetting rate $r^*$ can reach here, over all the possible values of the parameters $\alpha$ and $\beta$, as illustrated in Figs. \ref{fig:3}a-b.

\section{The regime $\alpha,\beta\gg r$ and the partial absorption problem}\label{secPartial}

We comment that the same expression (\ref{optBetaLarge}) was deduced in reference \cite{WhitehousePartialPRE2013} for diffusion under resetting with partial absorption: in that case, the dimensionless parameter $w$ was given by  $w=\sqrt{rD}/2\kappa$, where $\kappa$ is the absorption velocity of the target. 

The physical meaning of the approximation (\ref{mfptbeta}) can therefore be traced back to the problem of diffusion under resetting in the presence of a partially absorbing target\cite{WhitehousePartialPRE2013}. In that problem, a searcher performs diffusion with stochastic resetting to the initial position whereas a partially absorbing target is located at the origin. Upon target encounters, the searcher will not be necessarily absorbed at the target boundary but instead  reflected at some rate, such that the probability density $p(x,t)$ of the position $x$ will satisfy the so-called radiation boundary condition
\begin{equation}
   D \frac{\partial p(x,t)}{\partial y}\Big|_{x=0}=\kappa p(x=0,t),\label{rbc}
\end{equation}
where the absorption velocity $\kappa$ is the rate at which the searcher is absorbed at the target boundary. A different  interpretation of $\kappa$ can be found in Ref. \cite{schumm2021search} , where the searcher can diffuse inside the target, which is considered to have a certain thickness. In this configuration, $\kappa$ is proportional to the rate at which the searcher is absorbed while it is in the target region. 
Both interpretations lead to the same results when the target size tends to zero, which is the case of interest here.

It is found that the mean time at which the searcher reacts with the target is given by \cite{WhitehousePartialPRE2013}
\begin{equation}
    T_{p}(x_0)=\frac{e^{\sqrt{\frac{r}{D}}x_0}-1}{r}+\frac{e^{\sqrt{\frac{r}{D}}x_0}}{\kappa\sqrt{r/D}},\label{mfptpartial}
\end{equation}
where the other parameters $r$, $x_0$ and $D$ are the same as in our model.

By simple inspection, one can notice that Eq.  (\ref{mfptpartial}) has the same form as the approximation  (\ref{mfptbeta}) of $T_{av}$ in the limit of high transition rates ($\alpha,\beta$). Although the radiation boundary condition does not assume any internal target dynamics, we can make a mapping between the parameters $\alpha$ and $\beta$ and an absorption velocity $\kappa$ through the equation
\begin{equation}
    \kappa=\frac{\alpha}{\beta}\sqrt{\alpha+\beta}\sqrt{D}.\label{kappaequ}
\end{equation}
In other words, the optimal resetting rate in the problem of partial absorption is given by solving Eq. (\ref{optBetaLarge}) with $w=\sqrt{rD}/2\kappa$ \cite{WhitehousePartialPRE2013}. Therefore, the solution $r^*(\beta=\infty)=D/x_0^2$ of Eq. (\ref{roptlargebeta}) coincides with the optimal rate in the case of weak absorption, $\kappa\ll \sqrt{Dr}$ \cite{WhitehousePartialPRE2013,schumm2021search}. However, this mapping between the two models is not valid for intermediate values of the transition rates. With the radiation boundary condition (\ref{rbc}), the behaviour of the optimal resetting rate $r^*$ is monotonic with respect to the absorption velocity $\kappa$, whereas the gating dynamics on time-scales comparable or longer than the diffusion time give rise to a new non-monotonic behaviour with respect to the target reactivity (Fig. \ref{fig:3}).

These findings point out a close connection between partially absorbing and intermittent boundaries, a connection that has been revealed before in the context of simple diffusion \cite{PRLFirstHittingTimes} or run-and-tumble motion \cite{mercado2021first}. Eq. (\ref{kappaequ}) is independent of the resetting rate and actually coincides with the expression found in \cite{PRLFirstHittingTimes} for a free Brownian particle. Furthermore, in \cite{Lawley2015}, it was proved that the mean solution of the diffusion equation with a boundary condition switching infinitely fast between Dirichlet and Neumann conditions and with the boundary being in the Neumann condition most of the time, satisfies the Robin condition in a form equivalent to Eq. (\ref{kappaequ})  above. Similar homogenization methods have been applied for the solutions of parabolic
partial differential equations with intermittent boundaries \cite{Lawley2015Parabolic}.

\section{Coefficient of variation}\label{secCav}

In this section we analyze the coefficient of variation defined as $C_{av}=\langle (t-T_{av})^2\rangle/T_{av}^2$. This quantity represents the relative fluctuations of the first hitting time $t$, distributed according to the density $P_{av}(x,t)$, around its mean $T_{av}$. With the help of the relation (\ref{fhtd}), the coefficient of variation can be easily calculated:
\begin{equation}
    C_{av}=-\frac{2}{T_{av}^2 }\frac{\partial \widetilde{Q}_{av}(x_0,s)}{\partial s}\Big|_{s=0}-1.\label{coeffav}
\end{equation}

Given the expression of the survival probability $\widetilde{Q}_{av}(x_0,s)$ in Eq.  (\ref{Qavx0}), we can obtain the coefficient of variation in a straightforward manner after some algebraic manipulations. However, it is convenient here to rewrite Eq.  (\ref{Qavx0}) in terms of the survival probability $\widetilde{Q}_r(x_0,s)$ for the ungated case, given by\cite{Evans2011PRL}
\begin{equation}
\widetilde{Q}_r(x_0,s)=\frac{1-e^{-x_0\sqrt{\frac{s+r}{D}}}}{re^{-x_0\sqrt{\frac{s + r}{D}}}+s}.\label{Qungated}
\end{equation}
Let us introduce the function
\begin{equation}
    \widetilde{F}_r(x_0,s)=\frac{r e^{-\sqrt{\frac{s+r}{D}} x_0}+s}{\sqrt{s+r}}.
\end{equation}
With these definitions, the average survival probability is
\begin{equation}
    \widetilde{Q}_{av}(x_0,s)=\frac{\alpha \widetilde{F}_r(x_0,s)\widetilde{Q}_r(x_0,s) +\beta \widetilde{F}_r(x_0,s+\alpha+\beta)/(s+\alpha+\beta)}{\alpha \widetilde{F}_r(x_0,s)+s\beta \widetilde{F}_r(x_0,s+\alpha+\beta)/(s+\alpha+\beta)}.\label{QavF}
\end{equation}
After taking the derivative with respect to $s$, we obtain
\begin{eqnarray}
\frac{\partial \widetilde{Q}_{av}(x_0,s)}{\partial s}\Big|_{s=0}=&\frac{\partial \widetilde{Q}_r(x_0,s)}{\partial s}\Big|_{s=0}-\left[T_{av}(x_0)-T_r(x_0)\right]\left[T_{av}(x_0)+\frac{1}{\alpha+\beta}\right]\\
&+\frac{\beta}{\alpha (\alpha +\beta)}\frac{\partial}{\partial s} \left(\frac{\widetilde{F}_r(x_0,s+\alpha+\beta)}{\widetilde{F}_r(x_0,s)}\right)\Big|_{s=0},\nonumber\label{partialQav}
\end{eqnarray}
where $T_r(x_0)$ is the MFPT for the ungated case given by Eq.  (\ref{Tresetting}). From the above expression, $C_{av}$ is obtained in terms of the coefficient of variation $C_r$ in the ungated case, which is calculated from an equation equivalent to Eq.  (\ref{coeffav}), namely
\begin{equation}
    C_{r}=-\frac{2}{T_{r}^2 }\frac{\partial \widetilde{Q}_{r}(x_0,s)}{\partial s}\Big|_{s=0}-1.\label{coeffr}
\end{equation}
Substituting the partial derivative of $\widetilde{Q}_r(x_0,s)$ with respect to $s$ into Eq.  (\ref{partialQav}), one gets
\begin{eqnarray}
    C_{av}=&\left(\frac{T_r(x_0)}{T_{av}(x_0)}\right)^2\left(C_r+1\right)+2\left[1-\frac{T_r(x_0)}{T_{av}(x_0)}\right]\left[1+\frac{1}{(\alpha+\beta)T_{av}(x_0)}\right]\label{Cav}\\
    &-\frac{2\beta}{\alpha (\alpha +\beta)\left[T_{av}(x_0)\right]^2}\frac{\partial}{\partial s} \left(\frac{\widetilde{F}_r(x_0,s+\alpha+\beta)}{\widetilde{F}_r(x_0,s)}\right)\Big|_{s=0}-1\nonumber.
\end{eqnarray}

The advantage of expressing the coefficient of variation $C_{av}$ in terms of $C_r$ is to elucidate how different the fluctuations of the FHT for a dynamical target are from those of a simple target. Specially important to us is to see whether a generic feature of processes under resetting at the optimal rate also holds in our model. It is known that search processes under stochastic resetting which are optimal at a non-zero resetting rate, which is the case here, have a coefficient of variation equal to unity at optimality \cite{ReuveniPRLoptimal,PalPRL2017,BelanPRL2018}. This property holds true if the process is brought to the same initial state after each reset. In our case, this condition is not fulfilled, as resetting only acts on the particle and not on the target: after resetting the particle position, the target may not be in the state it occupied at $t=0$ (we compare in \ref{secBressloff} our results with the case where both the particle and the target are subject to resetting, as studied in \cite{Bressloff_2020}). In the following, we see that the aforementioned generic property holds in the limits $\beta\to0$ and $\beta\to\infty$, but is violated in the more general intermediate regime. 

\begin{figure}[htp]
    \centering
    \includegraphics[width=\textwidth]{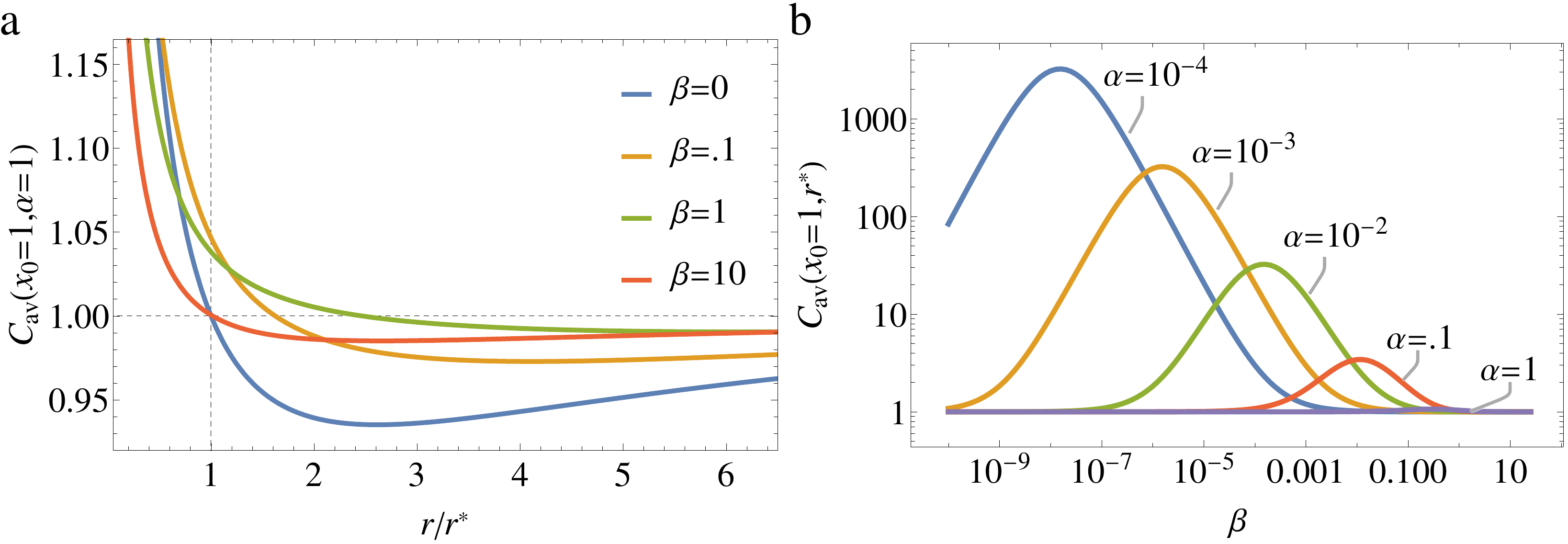}
    \caption{a) Coefficient of variation $C_{av}$ as a function of $r/r^{*}$ for fixed $x_0=1$, $D=1$, $\alpha=1$ and several values of $\beta$. b) Same quantity as a function of $\beta$ at the optimal rate $r^*$ (for $x_0=1$, $D=1$).}
    \label{fig:4ab}
\end{figure}

It is straightforward to notice that when $\beta=0$, we recover from Eq. (\ref{Cav}) the coefficient of variation for the ungated case, or  $C_{av}(\beta=0)=C_r$ [recall that  $T_{av}(\beta=0)=T_r$]. In the limit $\beta\to\infty$, the first hitting times diverges as $T_{av}\propto \sqrt{\beta}$ (see Eq.  \ref{mfptbeta}), and it is not difficult to see from the definition of $\widetilde{F}_r(x_0,s)$ that, in the limit of large $\beta$ and at the optimal resetting rate $r^*(\beta=\infty)=D/x_0^2$, 
\begin{equation}
    \frac{\partial}{\partial s} \left(\frac{\widetilde{F}_r(x_0,s+\alpha+\beta)}{\widetilde{F}_r(x_0,s)}\right)\Big|_{s=0}\propto \sqrt{\beta}.
\end{equation}
One deduces from Eq.  (\ref{Cav}) that $C_{av}(r^*,\beta\to\infty)\to 1$. These limiting behaviours are checked in Fig. \ref{fig:4ab}b with the exact solution. 

Whereas the relative fluctuations of the 
first hitting times are unity at optimality in the cases $\beta=0$ and $\beta=\infty$, this property is not general. The intricate way in which Eq.  (\ref{Cav}) depends on the target intermittency parameters does not allow an explicit analysis at finite $\alpha$ and $\beta$. Nevertheless, we performed a numerical evaluation of Eq.  (\ref{Cav}) in a wide range of values of $\beta$ and $\alpha$ at the corresponding optimal resetting rate $r^*$. The results are shown in Figs. \ref{fig:4ab}a-b, where the coefficient of variation takes values different from unity. As displayed in Fig. \ref{fig:4ab}b, when the target spends long periods of time in the two states, \textit{i.e.}, when $\alpha,\beta\ll D/x_0^2$, the quantity $C_{av}$ can take values much larger than $1$ at optimality, even when the target is reactive most of the time ($\beta\ll \alpha$).

\section{Conclusion}\label{secDiscussion}

We have studied the statistical properties of the first hitting time between a diffusing particle undergoing stochastic resetting to the initial position and a target that intermittently switches between a reactive and a non-reactive state. We have calculated the mean time it takes for the particle to hit the target for the first time in its reactive state, and have shown that this quantity can be minimized with respect to the resetting rate. This feature is also characteristic of many resetting processes with perfectly absorbing targets.

The MFHT increases due to the intermittent dynamics of the target. The minimal MFHT can thus be very high when the target is mostly non-reactive, which is intuitive since the task of searching an intermittent target is much more challenging.

We have found that when the target becomes highly intermittent, \textit{i.e.}, when the transitions between the reactive and the non-reactive state occur over a time-scale much smaller than the diffusion time, the model is equivalent to the problem of a partially absorbing target. In this case, we could establish a relationship between the target rates, the diffusion coefficient and the effective absorption velocity of the radiation boundary condition. Such equivalence between partially absorbing and dynamical boundaries has been observed in other search processes \cite{PRLFirstHittingTimes,mercado2021first,Lawley2015,Lawley2015Parabolic}, but it does not hold in general. For instance, when the target transition rates are comparable to the inverse diffusion time, the optimal resetting rate exhibits distinctive features, such as a non-monotonic behaviours. 

It is also worth noting that the coefficient of variation of the search time is not unity at optimality, in contrast with resetting problems that have a complete renewal structure. Here, the coefficient of variation can reach values much larger than one at the optimal resetting rate, specially for targets that spend a moderate fraction of time in the inactive state but long periods of time in each state. Other situations are analogous to the different resetting protocols of the stochastic gate. For instance, a run-and-tumble particle can be stochastically reset to its initial position, or may also have its velocity reset according to a given distribution \cite{Evans_2018}. In continuous time random walks, both the position and the waiting time may be subject to reset, or only the position \cite{KusmierzPRE2019}. The scaled Brownian motion model has also been studied under complete \cite{bodrova2019scaled} or incomplete  \cite{bodrova2019nonrenewal} resetting protocols. 

Our results highlight how target internal dynamics, a widely observed feature in natural systems, affect the optimisation of random searches by resetting. The scope of this work can be extended to the study of non-Poissonian resetting/target switching, as well as to anomalous diffusion processes. Although we have considered here the resetting of a single particle in the presence of an intermittent target, our results can be generalized to extended systems that can be reset to a specific configuration. An illustrative example is the growth of an interface which is stochastically interrupted by resetting to a certain profile, as occurring in mammalian tumors that are reduced to their initial size when a chemical is applied \cite{Gupta2014}. It would be interesting to study interface growth under resetting when the system is surrounded by fluctuating boundaries.

\ack 
GMV thanks CONACYT (Mexico) for a scholarship support. We acknowledge support from Ciencia de Frontera 2019 (CONACYT), project \lq\lq Sistemas complejos estoc\'asticos: Agentes m\'oviles, difusi\'on de part\'iculas, y din\'amica de espines\rq\rq \ (Grant 10872).

\appendix

\section{Backward Fokker-Planck equations}\label{appendixA}

In this section we derive Eqs. (\ref{BFPQ0}) and (\ref{BFPQ1}), for a particle  located at $x_0$ at $t=0$. Let us first suppose that the target is initially non-reactive. In a realization of the search process, during a small time interval $[0,\Delta t]$, with probability $\alpha\Delta t$ the target will switch to the reactive state, and with probability $1-\alpha\Delta t$ it will remain non-reactive. Meanwhile, with probability $r\Delta t$, the particle will reset to the position $x_r$ and with probability $1-r\Delta t$, it will diffuse and reach a new position $x_0+\xi$, where $\xi$ is a small random displacement due to Brownian diffusion during $\Delta t$. The position at $\Delta t$ ($x_r$ or $x_0+\xi$) is considered as a new starting position, from which the particle may survive during the interval $[\Delta t,t+\Delta t]$, which is of length $t$. Summing the contributions of the various eventualities, we obtain the evolution of the survival probability at $t+\Delta t$, starting from $x_0$:
\begin{eqnarray}
  \fl   Q_0(x_0,t+\Delta t)=&(1-r\Delta t)\left[\alpha \Delta t \int d\xi Q_1(x_0+\xi,t)P_{\Delta t}(\xi)+(1-\alpha \Delta t)\int d\xi Q_0(x_0+\xi,t)P_{\Delta t}(\xi)\right], \\
    &+r\Delta t\left[\alpha \Delta t Q_1(x_r,t)+(1-\alpha \Delta t) Q_0(x_r,t)\right]\nonumber\label{A1}
\end{eqnarray}
where $P_{\Delta t}(\xi)$ is the density of $\xi$.

We expand the survival probabilities in the right-hand-side in series of $\xi$, which is Gaussian distributed with first moment $\langle \xi\rangle=0$ and second moment $\langle \xi^2\rangle=2D\Delta t$, with $D$ the  diffusion coefficient. The integrals in Eq. (\ref{A1}) are $\langle Q(x_0+\xi,t)\rangle_{\xi}\approx Q(x_0,t)+D\Delta t\frac{\partial^2 Q(x_0,t)}{\partial x_0^2}$. Neglecting the terms of order higher than $\Delta t$, one obtains  
\begin{eqnarray}
  \fl   Q_0(x_0,t+\Delta t)=Q_0(x_0,t)+\Delta t\left\{D\frac{\partial^2 Q_0(x_0,t)}{\partial x_0^2}+\alpha Q_1(x_0,t)-(r+\alpha)Q_0(x_0,t)+r Q_0(x_r,t)\right\}.\label{BFPEQ0expand}
\end{eqnarray}
Similarly, for the initial target state $\sigma=1$, we have 
\begin{eqnarray}
 \fl    Q_1(x_0,t+\Delta t)=Q_1(x_0,t)+\Delta t\left\{D\frac{\partial^2 Q_1(x_0,t)}{\partial x_0^2}+\beta Q_0(x_0,t)-(r+\beta)Q_1(x_0,t)+rQ_1(x_r,t)\right\}.\label{BFPEQ1expand}
\end{eqnarray}
In the limit $\Delta t\to 0$, Eqs. (\ref{BFPEQ0expand}) and (\ref{BFPEQ0expand}) become (\ref{BFPQ0}) and (\ref{BFPQ1}), repectively.

\section{Comparison with the Bressloff's model}\label{secBressloff} 
In this section we compare our expression for the MFHT, Eq.  (\ref{mfptTav}), with the analogous quantity deduced by Bressloff in \cite{Bressloff_2020}. In this work, a one dimensional Brownian particle diffuses in the interval $[0,L]$ and is subject to stochastic resetting to the initial position $x_0$, with $0<x_0<L$. A dynamic target placed at the origin switches between an active absorbing state and a reflecting state which prevents absorption. The MFHT for this model is given by equation (4.19) in  \cite{Bressloff_2020}, from which we can obtain the MFHT in the semi-infinite domain by taking the limit $L\to\infty$:
\begin{eqnarray}
    T_{B}(x_0)&=\frac{e^{\sqrt{\frac{r}{D}}x_0}-1}{r}+\frac{\beta e^{\sqrt{\frac{r}{D}}x_0}}{\alpha\sqrt{r(r+\alpha+\beta)}},\label{mfhtBress}
\end{eqnarray}
with the same notation for the switching rates $\alpha$ and $\beta$ than ours.
Although the model studied in \cite{Bressloff_2020} is very similar, it bears an important difference. In \cite{Bressloff_2020}, when the particle is reset to $x_0$, the state of the target is also re-initialised to the state $\sigma=0$ [with probability $\beta/(\alpha+\beta)$] or $\sigma=1 $ [with probability $\alpha/(\alpha+\beta)$]. Conversely, in our model, the dynamics of the intermittent target is completely independent of the particle dynamics and not subject to resetting. This leads to quite different results for the behaviour of the mean time to absorption. 

\begin{figure}[htp]
    \centering
    \includegraphics[width=\textwidth]{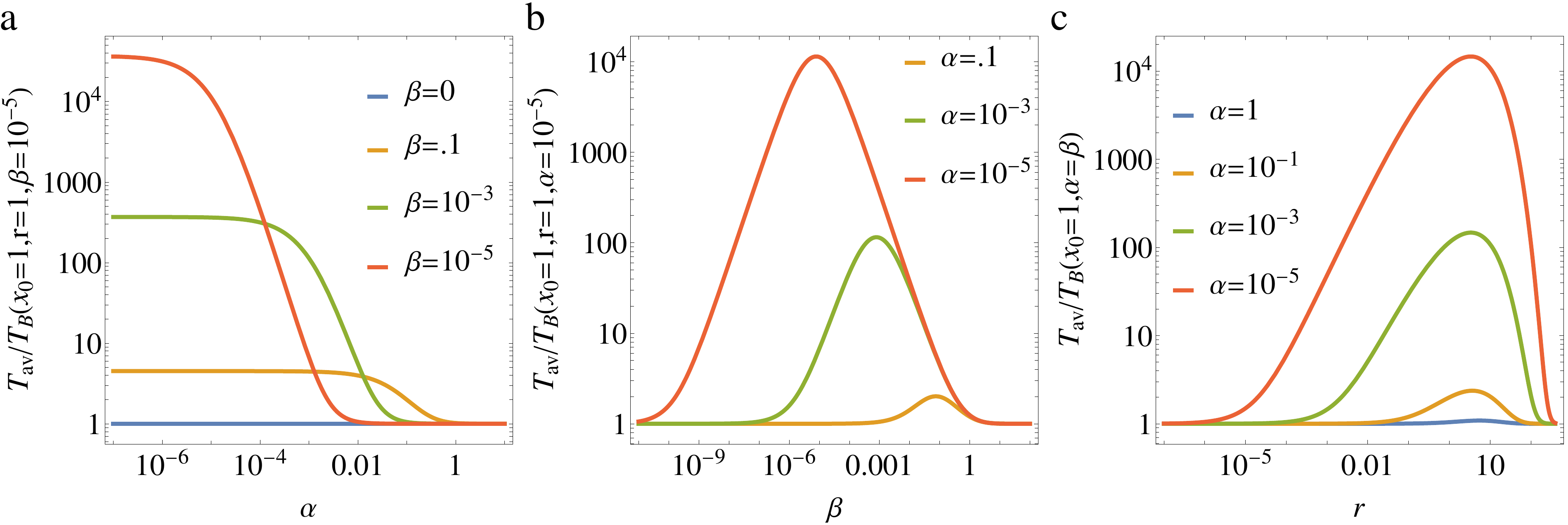}
    \caption{(a) Mean first hitting times $T_{av}/T_{B}$ as a function of $\alpha$ and several values of $\beta$ at $r=1$. (b) Same quantity as a function of $\beta$ for several values of $\alpha$. (c) $T_{av}/T_{B}$ as a function of $r$ for several values of $\alpha$ (and $\beta=\alpha$). In all cases, $x_0=1$ and $D=1$.}
    \label{fig:5ac}
\end{figure}

Eq.  (\ref{mfhtBress}) can be rewritten in terms of $T_{av}(x_0)$ here as
\begin{equation}
    T_{B}(x_0)=T_{av}(x_0)-\frac{\beta\sqrt{r} e^{\left(\sqrt{\frac{r}{D}}-\sqrt{\frac{r+\alpha+\beta}{D}}\right)x_0}}{\alpha(\alpha+\beta)\sqrt{r+\alpha+\beta}},\label{BvsTav}
\end{equation}
which implies that $T_{B}$ is lower than $T_{av}$ for all non-zero values of the parameters $\alpha$, $\beta$ and $r$. It is easy to notice that the difference between both quantities can become very large for cases in which $\alpha$ and  $\beta$ are $\ll r$ (see Figs. \ref{fig:5ac}a-c). 

To further contrast between these results, let us analyse the limiting case in which the particle resets to the origin ($x_0=0$) at infinite rate ($r=\infty$). In this scenario, once the search process has started the particle immediately returns to the origin, with the target still being in its initial state. If the target is initially in the reactive state ($\sigma=1$), the particle will be immediately absorbed, yielding to $T_1=0$. If the target is initially in the non-reactive state ($\sigma=0$), the particle will remain at the origin (due to the infinitely frequent resetting) until the target switches to the reactive state with rate $\alpha$, in this case $T_0=1/\alpha$. Therefore, from the definition of $T_{av}$, one obtains
\begin{equation}
    T_{av}(x_0=0,r=\infty)=\frac{\beta}{\alpha(\alpha+\beta)},
\end{equation}
which in fact coincides with Eq.  (\ref{mfptTav}). Conversely, from Eq.  (\ref{mfhtBress}) one can easily see that
\begin{equation}
    T_{B}(x_0=0,r=\infty)=0,
\end{equation}
\textit{i.e.}, in \cite{Bressloff_2020} the particle is immediately absorbed irrespective the initial target state. This is a consequence of the resetting process which, being infinitely frequent, makes the target rapidly active, even if $\frac{\alpha}{\alpha+\beta}\ll 1$. In this model, stochastic resetting enhances target detection not only by means of the particle motion but also by promoting target activation.

We notice in Fig. \ref{fig:5ac} that $T_{av}$ approaches $T_B$ in the limit of high switching rates, \textit{i.e.}, when $\alpha,\beta\gg r$. This can be seen directly from Eq.  (\ref{BvsTav}), where the second term of the right-hand-side approaches zero in this limit. Furthermore, when $\beta\to0$, the two solutions $T_{av}(x_0)$ and $T_B(x_0)$ tend to that of the ungated case, given by $T_r(x_0)$ in Eq. (\ref{Tresetting}).

\section*{References}
\bibliographystyle{iopart-num}
\bibliography{biblio}

\end{document}